\documentclass[twocolumn,prb,preprintnumbers,superscriptaddress,amsmath,amssymb,english]{revtex4}
\usepackage{amsmath,amssymb,amsfonts,mathrsfs}
\usepackage{amsthm}
\usepackage{CJK}
\usepackage{bm}
\usepackage{booktabs}
\usepackage{multirow}
\usepackage{longtable}
\usepackage{epstopdf}
\usepackage{dcolumn} 
\usepackage{babel}
\usepackage{graphicx}
\usepackage{float}
\allowdisplaybreaks
\usepackage{braket}
\usepackage[breaklinks]{hyperref}
\usepackage{color}
\hypersetup{colorlinks=true, linkcolor=blue, citecolor=blue, filecolor=blue, urlcolor=blue}

\begin{document}
    \title{Single pair of charge-two high-fold fermions with type-II van Hove singularities on the surface of ultralight chiral crystals}

    \author{Xiaoliang Xiao}
    \affiliation{Institute for Structure and Function $\&$ Department of Physics, Chongqing University, Chongqing 400044, China}   
    
    \author{Yuanjun Jin}
    \affiliation{Division of Physics and Applied Physics, School of Physical and Mathematical Sciences, Nanyang Technological University, Singapore 637371, Singapore}  
    
    \author{Da-Shuai Ma}
    \affiliation{Institute for Structure and Function $\&$ Department of Physics, Chongqing University, Chongqing 400044, China}
    \affiliation{Center of Quantum materials and devices, Chongqing University, Chongqing 400044, China} 
    
    \author{Haoran Wei}
    \affiliation{Institute for Structure and Function $\&$ Department of Physics, Chongqing University, Chongqing 400044, China}
    
    \author{Jing Fan}
    \affiliation{Center for Computational Science and Engineering, Southern University of Science and Technology, Shenzhen 518055, China} 
    
    \author{Rui Wang}
    \affiliation{Institute for Structure and Function $\&$ Department of Physics, Chongqing University, Chongqing 400044, China}
    \affiliation{Center of Quantum materials and devices, Chongqing University, Chongqing 400044, China}
    \affiliation{Chongqing Key Laboratory for Strongly Coupled Physics, Chongqing 400044, China}
    
    \author{Xiaozhi Wu}
    \email[]{xiaozhiwu@cqu.edu.cn}
    \affiliation{Institute for Structure and Function $\&$ Department of Physics, Chongqing University, Chongqing 400044, China}
    \affiliation{Chongqing Key Laboratory for Strongly Coupled Physics, Chongqing 400044, China}
    
    \begin{abstract}
        The realization of single-pair chiral fermions in Weyl systems remains challenging in topology physics, especially for the systems with higher chiral charges $C$. In this work, based on the symmetry analysis, low-energy effective model, and first-principles calculations, we identify the single-pair high-fold fermions in chiral cubic lattices. We first derive the minimal lattice model that exhibits a single pair of Weyl points with the opposite chiral charges of $C$ = $\pm{2}$.  Furthermore, we show the ultralight chiral crystal P4$_3$32-type LiC$_2$ and its mirror enantiomer as high-quality candidate materials, which exhibit large energy windows to surmount the interruption of irrelevant bands. Since two enantiomers are connected by the mirror symmetry, we observe the opposite chiral charges $C$ and the reversal of the Fermi arc velocities, showing the correspondence of chirality in the momentum space and the real space. In addition, we also reveal type-II van Hove singularities on the helicoid surfaces, which may induce chirality-locked charge density waves on the crystal surface. Our work not only provides a promising platform for controlling the sign of topological charge through the structural chirality but also facilitates the exploration of electronic correlations on the surface of ultralight chiral crystals.
    \end{abstract}
    
    
    
    \maketitle
    
    \textit{{\color{blue}Introduction. ---}}
    The classification of the band topology, by the crystallographic space group (SG) symmetry, can give rise to distinct types of unconventionally chiral Weyl fermions (CWFs) \cite{Un-1, Un-2, Un-3, Un-4, HWF-Review, Model-SG230}, which are usually characterized by larger chiral charges $C$ or higher degeneracies at crossing points. To date, they roughly contain twofold quadratic (cubic and quadruple) fermions with $C$ = $\pm$2 ($\pm$3 and $\pm$4) \cite{MWSM-C-123, CWSM-C-123, Jinyj, Un-2, two-band-C-4, HODFs, Model-C-4, C-4-cP-C24}, threefold charge-two (C-2) spin-1 fermions \cite{Un-1, Model-C-2-TDP}, fourfold C-2 Dirac and C-4 spin-$\frac{3} {2}$ fermions \cite{Un-1, Un-3, Un-4}, sixfold C-4 double spin-1 fermions \cite{Un-1}, and so on \cite{Model-SG230}. The chiral crystal \cite{Chiral-crystal}, lacking the inversion, mirror, and roto-inversion symmetry, is regarded as a natural platform for achieving these CWFs in condensed matter physics. Besides, exploring these unconventional CWFs with nontrivial chiral charges in chiral crystals can provide a new perspective for discovering interesting physical properties, such as chiral zero sound \cite{ZS, ZS-1}, longer Fermi arc surface states \cite{MTTSM-TMDS, RhSi-1, CoSi, CoSi-1, XSi, CTSMC4}, chiral/axial anomaly \cite{CA, CA-1, CA-2, CA-3, CA-4, CA-5}, larger quantized circular photogalvanic effect \cite{CPGE, CPGE-1, COR-MWSM}, chiral Landau levels/bands \cite{LL, LL-1, LL-2, LL-3}, and $etc$. Recently, the systems with single-pair Weyl points (WPs) have attracted significant attention \cite{Model-SP-C-2-SG230, Model-SP-C-1} since their topological surface states (TSSs) can be easily imaged in spectroscopy experiments \cite{Exp-C-4}. Interestingly, on the (001) surface of the XSi (X = Co, Rh) crystals, type-II van Hove singularities (VHSs) and chirality-locked charge density wave (CDW) were observed to manifest at the helicoid surface states \cite{sanchez2023tunable, li2022chirality, rao2023charge}. This discovery opens up possibilities for studying interactions-driven ordered states on the surface of topological materials, which may result in $p$ wave pairing superconductivity\cite{w57, Psc}.  So far, most of the research for such single-pair WPs is focused on the C-1, two-band, and bosonic systems \cite{Model-SP-C-1, Phonon-C-4-1, Phonon-C-4-2, Phonon-C-4-5, Model-SP-C-2-SG230, soin-SP-C-1, Model-SP-C-2}. However, the single-pair high-fold WPs, \textit{i.e.}, coexisting spin-1 Weyl point (SWP) and C-2 Dirac point (C-2 DP), with the opposite chirality of $C$ = $\pm{2}$ are rarely reported in fermionic systems. Therefore, it is eager to search for realistic materials to study such single-pair CWFs. 
    
\begin{figure}
        \setlength{\belowcaptionskip}{-0.1cm}
        \setlength{\abovecaptionskip}{0.05cm}
        \centering
        \includegraphics[scale=1.35]{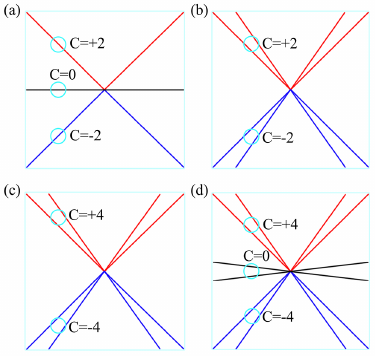}
        \caption{Four types of unconventionally chiral fermions with the high-fold degeneracy. (a) Threefold spin-1 Weyl fermions. (b) Fourfold charge-2 Dirac fermions, which are equivalent to adding two spin-$\frac {1} {2}$ Weyl fermions with the same chirality. (c) Fourfold spin-$\frac {3} {2}$ Weyl fermions. (d) Sixfold double spin-1 Weyl fermions, which are a composition of two identical spin-1 Weyl fermions. The bands marked the corresponding chiral charges, respectively.  
        \label{figure-0}}
    \end{figure}

    Until now, the experimental and theoretical realization of only several materials\textit{---}including XSi (X = Co, Rh), AlPt, PdBiSb, and PdGa crystals with SG P2$_1$3 [No. 198], namely P2$_1$3-type materials\textit{---}has been proposed \cite{MTTSM-TMDS, RhSi, RhSi-1, CoSi, CoSi-1, XSi, CTSMC4, CTSMC4-1, PdBiSe, PtGa}, while it, as a matter of fact, carries chiral charges of $C$ = $\pm{4}$ but $C$ = $\pm{2}$. This is because, in these material systems, a SWP splits into a fourfold spin-$\frac{3}{2}$ DP [Figs. \ref{figure-0}(a) and \ref{figure-0}(c)] and a C-2 DP evolves into a sixfold double spin-1 WP [Figs. \ref{figure-0}(b) and \ref{figure-0}(d)] under the non-negligible spin-orbit coupling (SOC) \cite{CoSi-C-2-SOC, CoGe-SOC}. Thus, it is mainly realized in bosonic and artificial systems \cite{DWP-TMDS, FeSi, Phonon-C-2, Phonon-C-2-DWSM-jin, CsBeF5}. Fortunately, there is a class of ultralight materials composed of light atoms with the negligible SOC. In fact, the investigation of the topological fermions involving the ultralight materials is quite important in fermionic systems \cite{fco-C6, B4N4, SW-57C, WSCN, FWSM, MTCs, N4Cl, C-4-cP-C24, T-Carbon-exp, T-Carbon-exp-1}, in which SU(2) spin-rotational symmetry is conserved \cite{SU2}. Hence, these ultralight materials provide a promising platform to search for a class of candidates with the corresponding SGs that permit the existence of single-pair high-fold CWFs.  

    In this work, based on symmetry arguments and the low-energy effective model, we identify that the single-pair high-fold CWFs with chiral charges of $C$ = $\pm{2}$ can exist in chiral cubic crystals. We further construct the minimal lattice model, which captures all the topological properties and generates visible double-helicoid Fermi arcs spanning the entire (001) surface Brillouin zone (BZ). Furthermore, by first-principles calculations, we take the ultralight crystal LiC$_2$ crystallized in SG P4$_3$32, called P4$_3$32-type LiC$_2$, and its mirror enantiomer as concrete examples to confirm the single-pair high-fold CWFs. These materials have three distinctions compared to previously reported P2$_1$3-type materials. The first distinction is that the SOC effect is indeed pretty small in P4$_3$32-type LiC$_2$. The second distinction is that the C-2 DP in P4$_3$32-type LiC$_2$ is described by a corepresentation equivalent to a four-dimensional (4D) single-valued irreducible representation (SVIR), but the DP in P2$_1$3-type materials is described by the time-reversal ($\mathcal{T}$) symmetric corepresentation formed by pairing two two-dimensional (2D) SVIRs. The third distinction is that P4$_3$32-type LiC$_2$ possess the mirror enantiomer P4$_1$32-type LiC$_2$ with different SG. Besides, P4$_3$32-type LiC$_2$ is an ideal semimetal with clear high-fold fermions near the Fermi level, unlike P2$_1$3-type PdGa enantiomers which are involved trivial bands and whose chiral fermions are far away from the Fermi level\cite{CTSMC4-1}. Moreover, the two enantiomers with opposite chiral charges $C$ exhibit the reversal of Fermi-arc velocities. Our work provides an ideal platform to explore the single-pair high-fold WPs with $C$ = $\pm{2}$ and control the sign of chiral charges by the handness of chiral crystals. Finally, we unveil type-II VHSs on the (001) surface of P4$_3$32-type LiC$_2$, offering a promising platform to explore electronic correlated phenomena such as chirality-locked CDW, unconventional superconductivity, and more.
    
    \textit{{\color{blue}Symmetry arguments and minimal lattice model. ---}} 
    First, we begin analyzing the symmetry conditions to search for a single pair of CWFs with the coexistence of SWP and C-2 DP in spinless systems. The symmetry argument goes as follows. For nonmagnetic systems with single-pair WPs, their WPs must be located at the time-reversal-invariant momentum (TRIM) points \cite{No-Go-1, No-Go-2, Model-SP-C-2-SG230}. Hence, only high-symmetry points (HSPs) of the TRIM are considered. First of all, considering the C-2 DP, the band crossing $\emph{\textbf{k}}$ point should possess two 2D SVIRs or a 4D SVIR \cite{kp-method}. By scanning through all the SVIRs of the little group at HSPs for 230 SGs \cite{kp-method}, six SGs can be found, \textit{i.e.}, Nos. 19 (R), 92(A), 96 (A), 198 (R), 212 (R), and 213 (R). Here, it can discover that the minimal SG of the C-2 DP is SG 19 with two screw rotational symmetries ($\widetilde{C}_{2x}$ = $\{C_{2x}|\frac{1}{2} \frac{1}{2} 0\}$ and $\widetilde{C}_{2y}$ = $\{C_{2y}|0 \frac{1}{2} \frac{1}{2}\}$) and $\mathcal{T}$-symmetry. Among six SGs, three SGs [Nos. 198 (P2$_1$3), 212 (P4$_3$32), and 213 (P4$_1$32)] belong to PG $T$ or $O$ \cite{kp-method, COR-MWSM, Phonon-C-2-DWSM-jin} and own a three-dimensional (3D) SVIR at the $\Gamma$ point. This indicates that single-pair CWFs with a C-2 DP and a SWP can only occur in the three SGs above. The symmetry requirements are listed in Table \ref{table1} and the low-energy effective $\emph{\textbf{k}}\cdot \emph{\textbf{p}}$ models for the $\Gamma$ and $\mathrm{R}$ points are applied to prove the SWP and C-2 DP [see more details in the Supplemental Material (SM) \cite{SM}]. Here, three bands of the SWP have the chiral charges of $+2$, 0, and $-2$, respectively [see Fig. \ref{figure-0}(a)], which corresponds to the chiral pseudospin-1 fermions \cite{Un-1}. The C-2 DP can be represented as the direct sum of two WPs with the same chiral charges of $C$ = $\pm1$ [Fig. \ref{figure-0}(b)], which then contributes to the DP with the double charge. Additionally, from Table \ref{table1}, we can find that the minimal symmetry case is SG 198. 

    \setlength{\tabcolsep}{5pt} 
    \renewcommand\arraystretch{1.4} 
    \begin{table}	
        \vspace{-0.1cm}
        \setlength{\belowcaptionskip}{0.1cm}
        \setlength{\abovecaptionskip}{0.1cm}
        \centering
        \caption{Three chiral SGs hosting the single-pair high-fold chiral fermions (\textit{i.e.}, SWP and C-2 DP) with negligible SOC effect. Location is the high-symmetry momentum $\emph{\textbf{k}}$. Type is the types of WPs. Generators denote the point-group parts of the generators of the little group at $\emph{\textbf{k}}$.}
        \label{table1}
        \begin{tabular}{cccc}
            \toprule
            \hline\hline
            SG & Location & Type & Generators \\
            \midrule
            \cline{1-4}
            \multirow{2}*{P2$_1$3} & $\Gamma$: (0, 0, 0) & SWP & $\widetilde{C}^{+}_{3,111}$, $\widetilde{C}_{2x}$, $\widetilde{C}_{2y}$, $\mathcal{T}$ \\
            & $\mathrm{R}$: ($\frac{1} {2}$, $\frac{1} {2}$, $\frac{1} {2}$) & C-2 DP & $\widetilde{C}^{+}_{3,111}$, $\widetilde{C}_{2x}$, $\widetilde{C}_{2y}$, $\mathcal{T}$ \\
            \multirow{2}*{P4$_3$32} & $\Gamma$: (0, 0, 0) & SWP & $\widetilde{C}^{+}_{3,111}$, $\widetilde{C}^{+}_{4z}$, $\widetilde{C}_{2x}$, $\widetilde{C}_{2y}$, $\mathcal{T}$ \\
            & $\mathrm{R}$: ($\frac{1} {2}$, $\frac{1} {2}$, $\frac{1} {2}$) & C-2 DP & $\widetilde{C}^{+}_{3,111}$, $\widetilde{C}^{+}_{4z}$, $\widetilde{C}_{2x}$, $\widetilde{C}_{2y}$, $\mathcal{T}$ \\
            \multirow{2}*{P4$_1$32} & $\Gamma$: (0, 0, 0) & SWP & $\widetilde{C}^{+}_{3,111}$, $\widetilde{C}^{+}_{4z}$, $\widetilde{C}_{2x}$, $\widetilde{C}_{2y}$, $\mathcal{T}$ \\
            & $\mathrm{R}$: ($\frac{1} {2}$, $\frac{1} {2}$, $\frac{1} {2}$) & C-2 DP & $\widetilde{C}^{+}_{3,111}$, $\widetilde{C}^{+}_{4z}$, $\widetilde{C}_{2x}$, $\widetilde{C}_{2y}$, $\mathcal{T}$ \\
            \bottomrule
            \hline\hline
        \end{tabular}
    \end{table}
    
    \begin{figure}
        \setlength{\belowcaptionskip}{-0.1cm}
        \setlength{\abovecaptionskip}{0.05cm}
        \centering
        \includegraphics[scale=1.35]{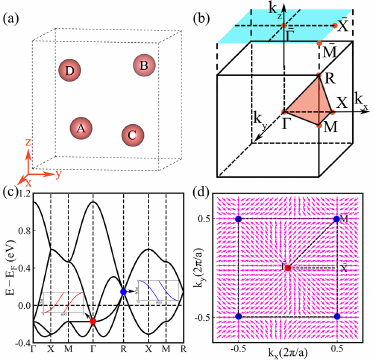} 
        \caption{The four-band minimal lattice model for SG 212. (a) The lattice structure of the 4a ($\{x, x, x\}$) Wyckoff position with $x$ = 0.125. (b) The bulk BZ and projected on the (001) surface BZ. (c) The bulk band structure of hopping parameters: $t_0$ = 0.145 and $t_1$ = 0.160. The left and right insets represent the evolutions of WCCs for the SWP and C-2 DP at the $\Gamma$ and R points, where $\varphi$ $\in$ [0, $\pi$] is the polar angle and $\theta$ $\in$ [0, 2 $\pi$] is the azimuthal angle in the spherical coordinate system. (d) The Berry curvature distributions on the (001) surface BZ. The SWP with $C$ = $+2$ as the “source” flows into the “sink” generated by the C-2 DP with $C$ = $-2$.  
        \label{figure-1}}
    \end{figure}
    
    Here, SGs 212 and 213 are the mirror enantiomers of each other, which are energy degenerate. Thus, in the main text, we only consider a four-band minimal lattice model with SG 212. Its symmetry operators include two twofold screw symmetries ($\widetilde{C}_{2x}$ = $\{C_{2x}|\frac{1}{2} \frac{1}{2} 0\}$ and $\widetilde{C}_{2y}$ = $\{C_{2y}|0 \frac{1}{2} \frac{1}{2}\}$), a threefold rotational symmetry ${C}^{+}_{3,111}$ along the [111] direction, a fourfold screw symmetry $\widetilde{C}^{+}_{4z}$ = $\{C_{4z}^{+}|\frac{3}{4} \frac{1}{4} \frac{3}{4}\}$, and $\mathcal{T}$-symmetry. Here, based on the band representations \cite{BR, BR-1, BR-2}, we select the 4a ($\{x, x, x\}$) Wyckoff position occupied by the \textit{s} orbital to construct this lattice model, as shown in Fig. \ref{figure-1}(a). It has the four equivalent positions $\{\{x, x, x\}, \{-x+\frac{1}{2}, -x, x+\frac{1}{2}\}, \{-x,x+\frac{1}{2}, -x+\frac{1}{2}\}, \{x+\frac{1}{2}, -x+\frac{1}{2}, -x\}\}$, where $x$ = 0.125, denoted by A, B, C, and D, respectively. Under the basis $\{\Phi_A(r), \Phi_B(r), \Phi_C(r), \Phi_D(r)\}$, the matrix representations of the symmetry operators are
    \begin{equation}\label{TB-SYM}
        \setlength{\belowdisplayskip}{5pt}
        \setlength{\abovedisplayskip}{5pt}
        \begin{split}
            &D(\widetilde{C}_{2x}) = \Gamma_{11}, \ \ \ \ \ \
            D(\widetilde{C}_{2y}) = \Gamma_{10}, \ \ \ \ \ \ 
            D(\mathcal{T}) = \Gamma_{00} \mathcal{K}, \\
            &D(C^{+}_{3,111}) = \left[
            \begin{array}{cccc}
            	1 & 0 & 0 & 0 \\
            	0 & 0 & 0 & 1 \\
            	0 & 1 & 0 & 0 \\
                    0 & 0 & 1 & 0 \\
            \end{array}
            \right], \ \ 
            D(\widetilde{C}^{+}_{4z}) = \left[
            \begin{array}{cccc}
            	0 & 0 & 0 & 1 \\
            	0 & 0 & 1 & 0 \\
            	1 & 0 & 0 & 0 \\
                    0 & 1 & 0 & 0 \\
            \end{array}
            \right],
        \end{split}
    \end{equation}
    where $\Gamma_{ij}$ = $\sigma_{i} \otimes \tau_{j}$ ($i, j = 0, 1, 2, 3$), $\sigma(\tau)_0$ is a $2 \times 2$ identity matrix, $\sigma(\tau)_i$ is Pauli matrix, and $\mathcal{K}$ is the complex conjugate operator. The Hamiltonian is restricted by these symmetries
    \begin{equation}\label{CON}
        \setlength{\belowdisplayskip}{4pt}
        \setlength{\abovedisplayskip}{4pt}
        \begin{split}	
            D(\mathcal{O}) \mathcal{H}(\boldsymbol{k}) D^{-1}(\mathcal{O}) = \mathcal{H}(\mathcal{O}\boldsymbol{k}),
        \end{split}
    \end{equation}
    where $\mathcal{O}$ stands for symmetry operators and it runs over all operators, and $D(\mathcal{O})$ is its matrix representation. Therefore, based on Eqs. (\ref{TB-SYM}) and (\ref{CON}), the effective lattice model with SG 212 in the spinless case reads \cite{tb-method, tb-method-1}
    \begin{equation}\label{TB-SG212}
        \setlength{\belowdisplayskip}{5pt}
        \setlength{\abovedisplayskip}{5pt}
        \begin{split}
            \mathcal{H}(\boldsymbol{k}) &= t_0 \Gamma_{00} + 2 t_1 \left[
            \begin{array}{cccc}
            	  0 & h_{12} & h_{13} & h_{14} \\
            	h_{12}^{*} & 0 & h_{23} & h_{24} \\
            	h_{13}^{*} & h_{23}^{*} & 0 & h_{34} \\
                    h_{14}^{*} & h_{24}^{*} & h_{34}^{*} & 0 \\
            \end{array}
            \right].
        \end{split}
    \end{equation}
    The matrix elements are listed as follows,
    \begin{equation}\label{TB-SG212-h}
        \setlength{\belowdisplayskip}{5pt}
        \setlength{\abovedisplayskip}{5pt}
        \begin{split}
            h_{12} &= \mathrm{cos}\frac{k_x}{2} (\mathrm{cos}\frac{k_y - k_z}{4} + i \mathrm{sin}\frac{k_y - k_z}{4}), \\
            h_{13} &= \mathrm{cos}\frac{k_y}{2} (\mathrm{cos}\frac{k_x - k_z}{4} + i \mathrm{sin}\frac{k_x - k_z}{4}), \\
            h_{14} &= \mathrm{cos}\frac{k_z}{2} (\mathrm{cos}\frac{k_x - k_y}{4} + i \mathrm{sin}\frac{k_x - k_y}{4}), \\
            h_{23} &= \mathrm{cos}\frac{k_z}{2} (\mathrm{cos}\frac{k_x + k_y}{4} + i \mathrm{sin}\frac{k_x + k_y}{4}), \\
            h_{24} &= \mathrm{cos}\frac{k_y}{2} (\mathrm{cos}\frac{k_x + k_z}{4} + i \mathrm{sin}\frac{k_x + k_z}{4}), \\
            h_{34} &= \mathrm{cos}\frac{k_x}{2} (\mathrm{cos}\frac{k_y + k_z}{4} + i \mathrm{sin}\frac{k_y + k_z}{4}).
        \end{split}
    \end{equation}
    Here, $t_{i}$ ($i = 0, 1$) denotes the hopping strength of considered couplings and is a real parameter. The topological properties of these WPs at the $\Gamma$ and $\mathrm{R}$ points are extremely stable and exhibit necessary degeneracy. Thus, we can randomly select these hopping parameters within a reasonable range and ensure that it occurs as a natural result. In the main text, we adopt $t_0$ = 0.145 and $t_1$ = 0.160 in Eq. (\ref{TB-SG212}). The four-band minimal lattice model for SG 198 is provided in the SM \cite{SM}. 

    \begin{figure}
        \setlength{\belowcaptionskip}{-0.1cm}
        \setlength{\abovecaptionskip}{0.05cm}
        \centering
        \includegraphics[scale=1.35]{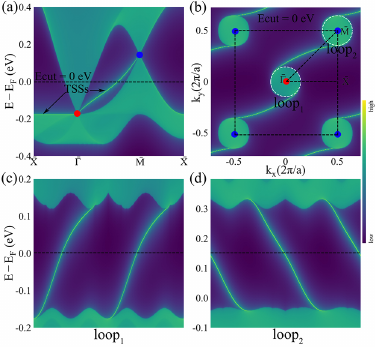}
        \caption{The TSSs and Fermi arcs of the four-band lattice model for SG 212. (a) The LDOS along $\bar{\mathrm{X}}$-$\bar{\mathrm{\Gamma}}$-$\bar{\mathrm{M}}$-$\bar{\mathrm{X}}$ paths on the (001) surface BZ. The black dashed line represents the isoenergy contour Ecut = $0$ eV. (b) The Fermi arcs at Ecut = $0$ eV and the associated WPs with the red and blue dots. Here, the red and blue dots indicate the positive and negative chiral charges. The double-helicoid surface arcs occur at $\bar{\Gamma}$ and $\bar{\mathrm{M}}$, respectively. (c) and (d) The surface LDOSs along the two white clockwise loops (loop$_1$ and loop$_2$) centered at $\bar{\Gamma}$ and $\bar{\mathrm{M}}$ in (b). The black dashed lines are the reference lines. It occurs in the opposite double-helicoid TSSs. 
        \label{figure-2}}
    \end{figure}
    
    The corresponding bulk BZ of the lattice structure of the lattice model is shown in Fig. \ref{figure-1}(b), where the cyan square represents the BZ projections in the (001) plane. The associated HSPs are also marked. Based on the hopping parameters above, we calculate the band dispersion along the high-symmetry lines (HSLs), as depicted in Fig. \ref{figure-1}(c), where a single pair of WPs do occur at the $\Gamma$ and R points (highlighted the red and blue dots). They have threefold and fourfold degeneracies with the linear dispersion, respectively. Then, to confirm the chiral charge of these WPs, we calculate the Wannier charge centers (WCCs) by using the Wilson-loop method \cite{WL}. We find that the threefold degenerate WP at the $\Gamma$ point possesses the chiral charge of $C$ = $+2$ [see the left panel of Fig. \ref{figure-1}(c)] and the chiral charge of the fourfold degenerate WP is $C$ = $-2$ at the R point, as shown in the right panel of Fig. \ref{figure-1}(c). This implies a SWP appears at $\Gamma$ and a C-2 DP at R. The total topological chiral charges are zero, which obeys the no-go theorem \cite{No-Go-1, No-Go-2}. Besides, the Berry curvature distribution of the (001) plane shows that there is only one ``source" at $\bar{\Gamma}$ and one ``sink" at $\bar{\mathrm{R}}$, and the SWP as the ``source" flows into the ``sink" generated by the C-2 DP [Fig. \ref{figure-1}(d)]. This is consistent with the associated chiral charges. These verify that the Hamiltonian can well describe the band topology of this kind of Weyl semimetal.
    
    \setlength{\tabcolsep}{5pt} 
    \renewcommand\arraystretch{1.4} 
    \begin{table*}	
        \vspace{-0.1cm}
        \setlength{\belowcaptionskip}{0.1cm}
        \setlength{\abovecaptionskip}{0.1cm}
        \centering
        \caption{The corresponding energies, positions, chiral charges, and multiplicities of the inequivalent WPs in LiC$_2$.}
        \label{table2}
        \begin{tabular}{cccccc}
            \toprule
            \hline\hline
            \ \ \ Material \ \ \ & \ \ \ \ WP \ \ \ \ & \ \ \ \ \ $E$-$E_\mathrm{F}$ (meV) \ \ \ \ & \ \ \ \ \ \ Coordinate ($k_1, k_2, k_3$) \ \ \ \ \ & \ \ \ \ Charge \ \ \ & \ \ \ Multiplicity \\ 
            \midrule
            \cline{1-6}
            \multirow{2}*{P4$_3$32-type LiC$_2$\ }
            & WP$_1$ & $31.29$ & (0, 0, 0) & $+2$ & 1 \\
            & WP$_2$ & $-202.11$ & (0.5, 0.5, 0.5) & $-2$ & 1 \\ 
            \bottomrule
            \hline\hline
        \end{tabular}
    \end{table*}

    \begin{figure}
        \setlength{\belowcaptionskip}{-0.1cm}
        \setlength{\abovecaptionskip}{0.05cm}
        \centering
        \includegraphics[scale=1.35]{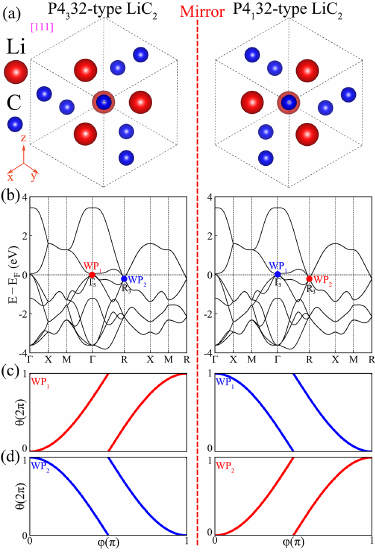}
        \caption{The representative materials for a single pair of WPs with the opposite chiral charges of $|C|$ = 2. (a) The views of the two enantiomers along the [111] direction, which are connected by the mirror symmetry. (b) The bulk band structures along the HSLs and the relevant SVIRs with the corresponding SGS P4$_3$32 and P4$_1$32. The band crossings near the Fermi level are marked WP$_1$ and WP$_2$, respectively. Here, the red and blue dots indicate the positive and negative chiral charges. (c) and (d) The evolutions of WCCs of the WPs for WP$_1$ and WP$_2$, respectively. The WPs of the two enantiomers are characterized by opposite chiral charges. 
        \label{figure-3}}
    \end{figure}
    
    The symmetry-guaranteed SWP and C-2 DP with opposite chiral charges can lead to unique nontrivial surface arcs due to the bulk-boundary correspondence. To confirm the topological signatures of the SWP and C-2 DP, we further calculate the local density of states (LDOSs) and Fermi arcs by employing the iterative Green’s function method \cite{Green1, Green2}. Note that, according to the crystal symmetry, (100), (010), and (001) surfaces are identical and support the same surface states. We here adopt a semi-infinite (001) surface. The $\Gamma$ and $\mathrm{R}$ points are projected to $\bar{\Gamma}$ and $\bar{\mathrm{M}}$ on the (001) surface BZ. As illustrated in Fig. \ref{figure-2}(a), two TSSs wind around the projected SWP and C-2 DP at $\bar{\Gamma}$ and $\bar{\mathrm{M}}$ [see black arrows]. In Fig. \ref{figure-2}(b), at the isoenergy contour Ecut = $0$ eV, the two Fermi arcs wind clockwise around the $\bar{\Gamma}$ point, while the two Fermi arcs wind anticlockwise around the $\bar{\mathrm{M}}$ point. This forms the double arc-shaped surface states that originate from the projection of the SWP and C-2 DP. In addition, it can find that two surface arcs are observed to link the projected SWP and C-2 DP because there is only a pair of WPs over the whole BZ, unlike the conventional WPs with opposite chiral charges exhibiting internally linked performance. This is consistent with the Berry curvature distribution. Moreover, to visualize the helicoidal surface states, we calculate the surface LDOSs along two clockwise loops (loop$_1$ and loop$_2$) centered at $\bar{\Gamma}$ and $\bar{\mathrm{M}}$ [Figs. \ref{figure-2}(c) and \ref{figure-2}(d)]. For the loop$_1$, two right-moving chiral edge modes appear inside the band gap. But the loop$_2$ encompasses the two left-moving chiral edge modes. This also hints they have opposite chirality with $C = \pm{2}$. 
    
    \textit{{\color{blue}Material realization with abundant real candidates. ---}}
    Symmetry arguments, low-energy effective models, and the minimal lattice models provide a fundamental theoretical strategy to investigate a single pair of WPs with the coexistence of the SWP and C-2 DP. Herein, by first-principles calculations, we take the realistic crystal material P4$_3$32-type LiC$_2$ and its mirror enantiomer P4$_1$32-type LiC$_2$, which are connected by the mirror symmetry [see Fig. \ref{figure-3}(a)], as examples to comfirm the chiral fermions with $C$ = $\pm{2}$. The other material candidates are summarized in the SM \cite{SM}. Top and side views of the crystal structure of P4$_3$32-type LiC$_2$ are shown in Fig. S2 in the SM \cite{SM}, which crystallizes in a chiral cubic lattice with SG P4$_3$32. Each unit cell contains four Li and twelve C atoms, occupying the 4a (0.125, 0.125, 0.125) and 8c (0.405, 0.405, 0.405) Wyckoff positions, denoted by red and blue balls, respectively. Its enantiomer with SG P4$_1$32 is displayed in Fig. S2 in the SM \cite{SM}, in which the Li and C atoms occupy two types of Wyckoff positions: 4a (0.875, 0.875, 0.875) and 8c (0.595, 0.595, 0.595). Their optimized lattice parameter is $|\mathbf{a}|$ = 4.31 {\AA}. They are mechanically, thermodynamically, and dynamically stable, based on the results of the elastic constants, the density-functional-base \textit{ab initio} molecular dynamics (AIMD) simulations, and the phonon dispersion along the HSLs [see Fig. S3 in the SM \cite{SM}]. 

    \begin{figure}
        \setlength{\belowcaptionskip}{-0.1cm}
        \setlength{\abovecaptionskip}{0.05cm}
        \centering
        \includegraphics[scale=1.35]{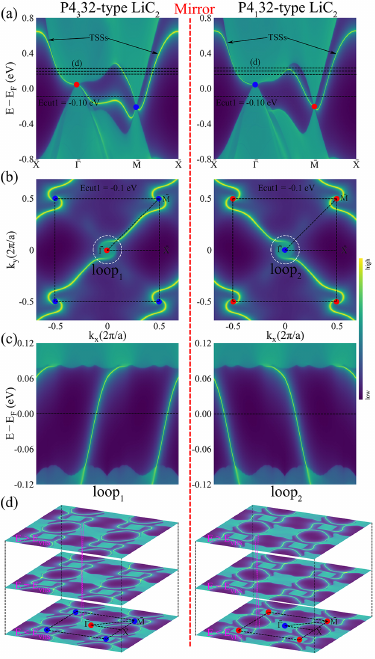}
        \caption{(a) The LDOSs of the two enantiomers along $\bar{\mathrm{X}}$-$\bar{\mathrm{\Gamma}}$-$\bar{\mathrm{M}}$-$\bar{\mathrm{X}}$ paths on the (001) surface BZ. The black dashed line represents different isoenergy contours. (b) Their Fermi arcs at Ecut1 = $-0.1$ eV and positions of the WPs. The double-helicoid surface arcs occur at $\bar{\Gamma}$ and $\bar{\mathrm{M}}$, respectively. (c) The surface LDOSs along the two white clockwise loops (loop$_1$ and loop$_2$) centered at $\bar{\Gamma}$ in (b). Two right-moving and left-moving chiral double-helicoid TSSs occur naturally. (d) Three isoenergy contours at three energies (0.20, 0.215, and 0.23 eV) indicated by three horizontal dashed lines in (a). Lifshitz transitions occur in the evolutions of their surface arc with the opposite position.
        \label{figure-4}}
    \end{figure}
    
    We next discuss the electronic properties of the two enantiomers. Since the Li atom is lighter than the C atom, they both are ideal spinless systems with the negligible SOC effect, as shown in Fig. S3 in the SM \cite{SM}. Hence, we only consider their band structures without SOC and mark the relevant SVIRs [Fig. \ref{figure-3}(b)]. The related symbols of SVIRs are listed in Table SIII-SIV in the SM \cite{SM}. We note that the two enantiomers are indeed energy degenerate due to the mirror symmetry. Besides, one shows clearly that the threefold degenerate WP$_1$ (SWP) with a 3D SVIR $\Gamma_5$ is located at $\mathrm{\Gamma}$ and the fourfold degenerate WP$_2$ (C-2 DP) with a 4D SVIR R$_3$, different from P2$_1$3-type crystals with two 2D SVIRs, is at the $\mathrm{R}$ point \cite{irvsp}. By carefully screening energy differences between the lowest conduction and the highest valence bands, we find only one pair of WPs in the whole BZ and verify the absence of any additional quasiparticles. This is consistent with the symmetry analysis and the lattice model. Then, to examine the chiral charges of different types of WPs, we employ the tight-binding (TB) Hamiltonian based on the maximally localized Wannier functions (MLWFs) \cite{Wannier90, Wannier90-1} to calculate the WCCs by using the Wilson-loop method \cite{WL}. As shown in Figs. \ref{figure-3}(c) and \ref{figure-3}(d), the evolutions of WCCs of P4$_3$32-type LiC$_2$ show that the WPs (SWP and C-2 DP) possess the chiral charges of $+2$ and $-2$, respectively. On the contrary, since Berry curvature is an axial tensor, its enantiomer P4$_1$32-type LiC$_2$ has the chiral charges of $-2$ and $+2$ at $\Gamma$ and R, respectively. This indicates that the chirality of chiral fermions can be reversed by altering the structural chirality, establishing a mapping between fermionic chirality and structural chirality. The sum of the chiral charges over the whole BZ both obeys the no-go theorem \cite{No-Go-1, No-Go-2}. The detailed information, including the corresponding energies, positions, chiral charges, and multiplicities, is listed in Table \ref{table2}. Moreover, it can find the SWP and C-2 DP, near the Fermi level, which can be viewed as ideal chiral fermions due to the ultraclean band dispersions. 
    
    To respect the chiral charges of the SWP and C-2 DP in the two enantiomers, we further calculate the LDOS and projected Fermi arcs by employing the WannierTools package \cite{WannierTools}. The LDOSs of the semi-infinite (001) surface of the two enantiomers are displayed in Fig. \ref{figure-4}(a). Two visible TSSs link the projected SWP and C-2 DP surrounding the projection points $\bar{\Gamma}$ and $\bar{\mathrm{M}}$. The alternating connection between starting and ending points leads to the Fermi arc traversing the entire BZ. The isoenergy contour of the (001) surface at Ecut = $-0.1$ eV provides the ultralong surface arcs in Fig. \ref{figure-4}(b). We can see that the Fermi arcs of two enantiomers possess a reversal of Fermi-arc velocities. Furthermore, we visualize the helicoidal (or double arc-shaped) surface arcs by calculating the surface LDOSs along two white clockwise loops (loop$_1$ and loop$_2$) centered at $\bar{\Gamma}$. It appears two right-moving chiral edge modes around $\bar{\Gamma}$ in P4$_3$32-type LiC$_2$ [see the left panel of Fig. \ref{figure-4}(c)]. As expected, owing to the mirror symmetry, the opposite situation is observed in its mirror enantiomer, as shown in the right panel of Fig. \ref{figure-4}(c). Besides, the evolutions of their surface arcs are illustrated by three isoenergy contours [Fig. \ref{figure-4}(d)], in which the surface arcs are visualized and their Lifshitz transitions at generic momenta, termed as type-II VHSs, occur clearly in the arc-like surface states with the opposite position. Such VHSs indicate electronic instability and may result in electronic correlations.  All topological properties of two enantiomers are consistent with the minimal lattice model. It is worth noting that they both host the ultralong Fermi arcs spanning the entire BZ, and their TSSs are not covered by the bulk band projection, which provides great promise for experimental detection and further applications.
    
    \textit{{\color{blue}Summary. ---}}
    In summary, based on symmetry arguments, low-energy effective models, and the minimal lattice model, we identify that three chiral SGs (Nos. 198, 212, and 213) can hold single-pair high-fold CWFs with $C$ = $\pm{2}$. Then, by first-principles calculations, we take the two enantiomers (P4$_3$32-type LiC$_2$ and P4$_1$32-type LiC$_2$) with high thermal stability as ideal examples to confirm it. They do appear ultralong visible double-helicoid Fermi arcs on the (001) surface BZ and own reversal of Fermi-arc velocities. The type-II VHSs on the (001) surface of P4$_1$32-type LiC$_2$ make it possible to promise an excellent playground for further observations of interactions-driven ordered states, such as the chirality-locked CDW, and superconductivity \cite{w57, Psc}. Besides, the CWFs also carry a monopole-like orbital-momentum locking texture, which may lead to a large orbital Hall effect, giant chirality-locked orbital magnetoelectric effect \cite{ome}.
    
    \textit{{\color{blue}Acknowledgement. ---}}
    This work was supported by the National Natural Science Foundation of China (NSFC, Grant No. 12174040, No.~12204074, No.~12222402, No.~11974062, and No. 12147102),  the China National Postdoctoral Program for Innovative Talent (Grant No. BX20220367), and Chongqing Natural Science Foundation (Grant No. cstc2020jcyj-msxmX0118).
    
    
    %
	
\end{document}